\documentclass[aps,preprint,a4paper,nofootinbib,showkeys]{revtex4}

\usepackage{xcolor}
\usepackage{amssymb,amsmath}
\usepackage{physics}
\usepackage{graphicx}

\usepackage{derivative}
\usepackage{hyperref}

\begin{document}

\title{A nonlinear Newtonian approximation to General Relativity}

\author{Oliver F. Piattella}
\email{of.piattella@uninsubria.it}
\affiliation{DiSAT, Universit\`a degli Studi dell'Insubria, via Valleggio 11, Como, Italy} 
\affiliation{INFN, Sezione di Milano, via Celoria 16, 20133, Milano, Italy}

\date{\today}

\begin{abstract}
We investigate a nonrelativistic approximation to general relativity that does not require the metric coefficients to be weak perturbations of the Minkowski metric. Instead, we assume a differential hierarchy in which the contributions to the curvature containing second derivatives of the metric dominate over those quadratic in its first derivatives, while the metric-dependent coefficients multiplying the retained terms are kept fully nonlinear. Since this separation is not covariant, we formulate the approximation in harmonic coordinates, where it corresponds to retaining the principal part of the reduced Einstein equations. We show that the truncated Einstein tensor satisfies the contracted Bianchi identities up to the order neglected in the derivative hierarchy, ensuring compatibility with energy--momentum conservation at the retained order. The standard Newtonian equations
are recovered in the weak-field regime, whereas outside it the dynamics retains a nonlinear dependence on the lapse and spatial
metric. In particular, null geodesics probe the spatial geometry at leading order, suggesting gravitational lensing as a natural setting in which observable departures from the standard weak-field description may arise. We illustrate the construction with a static spatially isotropic configuration and discuss the scope and limitations of the approximation.
\end{abstract}


\maketitle

\section{Introduction}

Newtonian gravity is described by the Poisson equation,
\begin{align}
    \nabla^2\Psi = 4\pi G\rho\,,
\end{align}
where $\Psi$ is the gravitational potential, $\rho$ is the mass density, and $G$ is Newton's gravitational constant, together with the equation of motion of a test particle,
\begin{align}
    \mathbf{a} = -\grad\Psi\,.
\end{align}
The value of the potential itself does not enter these equations: adding a constant to $\Psi$ leaves both the field equation and the particle dynamics unchanged. What determines the gravitational acceleration is the spatial variation of the potential.

In general relativity, the gravitational field is described by the metric, test particles follow geodesics, and the Einstein equations replace the Poisson equation. Unlike the Newtonian field equation, the Einstein equations depend explicitly on the gravitational variables as well as on their derivatives. In particular, the Ricci tensor contains terms of the schematic form
\begin{align}
    g^{-1}\partial^2 g\,,
    \qquad
    g^{-1}(\partial g)g^{-1}(\partial g)\,,
\end{align}
where $g^{-1}$ denotes the inverse metric, and indices have been suppressed.

This structural difference must, however, be interpreted with care. The components of the metric are coordinate dependent, and their apparent freedom includes gauge degrees of freedom. Consequently, the explicit occurrence of the metric in the field equations does not by itself establish a physical dependence on the values of its components. For example, a constant Lorentzian metric describes flat spacetime even when its components differ substantially from the standard Minkowski values, since these can be recovered by a linear change of coordinates. Any proposed departure from Newtonian dynamics must therefore be distinguished from effects arising solely from a choice of coordinates.

The usual weak-field derivation of Newtonian gravity starts from small perturbations of the Minkowski metric \cite{poisson2014gravity}. At leading order, the inverse metric multiplying derivatives of the perturbations can then be replaced by its Minkowski value, while terms quadratic in the perturbations are neglected. Together with an appropriate nonrelativistic ordering of the matter variables, this procedure recovers the Poisson equation and Newtonian particle dynamics as $c\to\infty$. The weak-field assumption and the nonrelativistic limit are, nevertheless, distinct ingredients of the derivation.

The relationship between general relativity and Newtonian gravity has also been studied in formulations that go beyond the elementary weak-field derivation. In the geometric approach developed by Trautman, Dautcourt, K\"unzle, and Ehlers, Newtonian gravity is formulated in terms of Newton--Cartan geometry, while Ehlers' frame theory provides a common framework in which relativistic and Galilean spacetime structures can be treated as members of a one-parameter family \cite{Ehlers:2019aco, Buchert:2019zup, Hartong:2022lsy}. Rigorous convergence results have also been established for particular matter models, including collisionless matter and perfect fluids \cite{Rendall:1993iz, Oliynyk:2007uno}. Starting from Newton--Cartan theory, systematic post-Newtonian corrections can be constructed and shown, under appropriate assumptions, to reproduce the usual post-Newtonian equations \cite{Dautcourt:1996pm}.

More recent work has shown that a nonrelativistic expansion of general relativity does not necessarily require weak gravitational fields. In a covariant large-$c$ expansion, relaxing the usual regularity assumptions on the relativistic connection allows a nontrivial lapse and strong gravitational time-dilation effects to survive in the nonrelativistic geometry \cite{VandenBleeken:2017rij}. This leads naturally to torsional Newton--Cartan structures and to a nonrelativistic gravitational theory extending ordinary Newtonian gravity. Action principles, matter couplings, and the associated gauge structure have subsequently been developed in Refs.~\cite{Hansen:2019vqf, Hansen:2020pqs}, while a complementary $3+1$ formulation makes explicit the additional strong gravitational potentials contained in the $1/c$ expansion \cite{Elbistan:2022plu}. A comprehensive review of these developments is given in Ref.~\cite{Hartong:2022lsy}.

The approximation investigated here is conceptually different. We do not construct a systematic expansion of the metric in powers of $1/c$, nor do we take a Newton--Cartan geometry as the leading-order spacetime structure. Instead, in a specified coordinate system, we assume a differential hierarchy in which the terms in the curvature containing second derivatives of the metric dominate over those quadratic in its first derivatives,
\begin{align}
    g^{-1}\partial^2 g
    \gg
    g^{-1}(\partial g)g^{-1}(\partial g)\,,
\end{align}
while the metric-dependent coefficients multiplying the second derivatives are retained without a weak-field expansion. The resulting equations, therefore, retain a nonlinear dependence on the metric itself.

This hierarchy is not manifestly invariant under general coordinate transformations, since the two structures displayed above are not separately tensorial. It should therefore be regarded, at this stage, as an approximation defined in a chosen coordinate system rather than as a covariant limit of general relativity. An essential part of the present analysis is to determine under which coordinate conditions and orderings this truncation is self-consistent, whether it remains compatible with the differential identities of the Einstein equations and with energy--momentum conservation, and whether it can lead to coordinate-independent observable effects.

A conceptually different correspondence between Newtonian gravity and general relativity has recently been discussed by Buchert \cite{Buchert:2023lxz}, who relates the Lagrangian equations of Newtonian irrotational dust to the Einstein equations by relaxing the integrability of the deformation coframes, without invoking either a weak-field expansion or a large-$c$ limit.

\section{The Newtonian limit: weak fields and large \texorpdfstring{$c$}{c}}\label{Sec:Newtlim}

We begin by briefly reviewing the standard derivation of Newtonian gravity from general relativity, in order to identify the assumptions that will be modified in the following section. The first is the weak-field approximation.

In a suitable coordinate system, we consider a one-parameter family of metrics of the form
\begin{align}
    g_{\mu\nu}(\epsilon)
    =
    \eta_{\mu\nu}
    +
    \epsilon h_{\mu\nu}
    +
    O(\epsilon^2)\,,
    \label{metexpansion}
\end{align}
where $\eta_{\mu\nu}$ is the Minkowski metric with a mostly positive signature, and $\epsilon$ is an expansion parameter controlling the departure from flat spacetime. We assume that $h_{\mu\nu}$ and its derivatives remain finite as $\epsilon\to0$.

At first order in $\epsilon$, the Einstein equations are linear in $h_{\mu\nu}$ and their second derivatives. The corresponding component equations are reported in Appendix~\ref{app:newtonian-details}. Since the zeroth-order geometry is Minkowski spacetime, the matter sources must be of the same perturbative order as the metric perturbation.

The weak-field approximation must be supplemented by a nonrelativistic ordering. We use $c$ to organize the expansion in dimensionless ratios and assume that the characteristic matter velocities and thermodynamic quantities remain finite, so that
\begin{align}
    \frac{v}{c}\to 0\,,
    \qquad
    \frac{p}{\rho c^2}\to 0\,, \qquad (c \to \infty)\,.
\end{align}
Here, $\rho$ denotes the rest-mass density, $p$ the isotropic pressure, $u^\mu$ the fluid four-velocity, and $v^i = dx^i/dt$ the ordinary coordinate three-velocity. We also denote by $T\equiv g^{\mu\nu}T_{\mu\nu}$ the trace of the energy--momentum tensor.

For a perfect fluid,
\begin{align}
    T_{\mu\nu}
    =
    \left(\rho+\frac{p}{c^2}\right)u_\mu u_\nu
    +
    pg_{\mu\nu}\,,
\end{align}
the leading large-$c$ behavior is
\begin{align}
    T_{00}
    &=
    \rho c^2+O(c^0)\,,
    &
    T_{0i}
    &=
    -\rho c v_i+O(c^{-1})\,,
    \\
    T_{ij}
    &=
    \rho v_i v_j+p\delta_{ij}+O(c^{-2})\,,
    &
    T
    &=
    -\rho c^2+O(c^0)\,.
\end{align}
Balancing the $00$ Einstein equation against the leading rest-mass source shows that the metric perturbation relevant for the Newtonian limit is of order $c^{-2}$. We may therefore absorb a fixed velocity scale into the definition of $h_{\mu\nu}$ and write
\begin{align}
    g_{\mu\nu}
    =
    \eta_{\mu\nu}
    +
    \frac{1}{c^2}h_{\mu\nu}
    +
    o(c^{-2})\,,
    \label{metexpansionc2}
\end{align}
so that $h_{\mu\nu}$ has dimensions of squared velocity.

Since $x^0 = ct$, each derivative with respect to $x^0$ introduces an additional inverse power of $c$ relative to a derivative with respect to $t$. At leading order in the large-$c$ expansion, the linearized Einstein equations reduce to
\begin{align}
\label{LinEE00cinftylim}
\nabla^2 h_{00} &= -8\pi G\rho\,,\\
\label{LinEE0icinftylim}
\nabla^2h_{0i} - h_{0k,ik} &= 0\,,\\
\label{LinEEijcinftylim}
h_{00,ij} - \nabla^2h_{ij} + h_{k(i,j)k} - h_{kk,ij}
&= 8\pi G\rho\,\delta_{ij}\,.
\end{align}
Hereafter, a comma followed by an index denotes partial differentiation with respect to the corresponding coordinate. Moreover, parentheses around spatial indices denote symmetrization without a factor $1/2$.

In the harmonic gauge, the leading-order gauge conditions are
\begin{align}
    h_{0k,k}=0\,,
    \qquad
    2h_{ik,k}
    =
    -h_{00,i}+h_{kk,i}\,,
\end{align}
and the field equations become
\begin{align}
    \nabla^2 h_{00}
    &=
    -8\pi G\rho\,,
    \\
    \nabla^2h_{0i}
    &=
    0\,,
    \\
    \nabla^2h_{ij}
    &=
    -8\pi G\rho\,\delta_{ij}\,.
\end{align}
It follows that
\begin{align}
    h_{ij}
    =
    h_{00}\delta_{ij}
    +
    H_{ij}\,,
    \qquad
    \nabla^2H_{ij}=0\,,
\end{align}
where $H_{ij}$ denotes the homogeneous part of the spatial metric perturbation. For a regular isolated system with the usual boundary conditions at spatial infinity, the homogeneous contribution can be eliminated, yielding
\begin{align}
    h_{ij}=h_{00}\delta_{ij}\,.
    \label{NewtonianSpatialMetric}
\end{align}
The homogeneous $0i$ sector can similarly be characterized by the Coriolis field
\begin{align}
    \omega_i
    =
    \varepsilon_{ijk}{h}_{0k,j}\,.
\end{align}
For regular isolated configurations with suitable boundary conditions at spatial infinity, this sector may be set to zero. More general boundary conditions or non-isolated configurations, however, may allow for non-vanishing Coriolis fields \cite{Ehlers:2019aco, Re:2025yms, Bianchi:2025rgi}.

The leading-order continuity and Euler equations are
\begin{align}
    \dot{\rho}
    +
    (\rho v_i)_{,i}
    &=
    0\,,
    \label{Continuitycinf}
    \\
    (\rho v_i)^{\bullet}
    +
    \left(
        \rho v_i v_j
        +
        p\delta_{ij}
    \right)_{,j}
    -
    \frac{1}{2}\rho h_{00,i}
    &=
    0\,.
    \label{Eulercinf}
\end{align}
Likewise, for a slowly moving massive test particle, the geodesic equation reduces to
\begin{align}
    \frac{d^2x^i}{dt^2}
    =
    \frac{1}{2}h_{00,i}\,.
    \label{NewtonianGeodesic}
\end{align}
Defining
\begin{align}
    h_{00} = -2\Psi\,,
\end{align}
the standard Newtonian theory is recovered in its usual formulation.

The spatial metric perturbation does not contribute to the leading-order motion of slowly moving massive particles, but it is nevertheless part of the relativistic gravitational field. In particular, it contributes at the same order as $g_{00}$ to null
geodesics and accounts for the standard relativistic correction to light deflection. The explicit derivation, together with the complete linearized field, conservation, and geodesic equations, is given in Appendix~\ref{app:newtonian-details}.

Note that the large-$c$ ordering does not imply that the gravitational field is static. The matter variables and the Newtonian potential may depend on time; what becomes subleading at Newtonian order are the  relativistic propagation terms containing time derivatives in the gravitational field equations.

\section{A derivative-truncated nonrelativistic approximation}
\label{sec:nonlinear}

We now investigate whether the weak-field restriction on the metric coefficients can be relaxed while retaining a nonrelativistic ordering of the matter variables. In contrast with the standard derivation reviewed in Sec.~\ref{Sec:Newtlim}, we do not assume that the metric is a small perturbation of the Minkowski metric. The components $g_{\mu\nu}$ and their inverse $g^{\mu\nu}$ are instead retained without expansion.

The Ricci tensor can be separated schematically into two contributions,
\begin{align}
    R_{\mu\nu}
    =
    R^{(2)}_{\mu\nu}
    +
    R^{(1,1)}_{\mu\nu}\,,
\end{align}
where $R^{(2)}_{\mu\nu}$ contains terms linear in the second derivatives of the metric,
\begin{align}
    R^{(2)}
    \sim
    g^{-1}\partial^2 g\,,
\end{align}
whereas $R^{(1,1)}_{\mu\nu}$ contains terms that are quadratic in the metric first derivatives,
\begin{align}
    R^{(1,1)}
    \sim
    g^{-1}(\partial g)g^{-1}(\partial g)\,.
\end{align}
The approximation considered here consists of assuming, in a specified coordinate system, the hierarchy
\begin{align}
    \left|
        R^{(1,1)}_{\mu\nu}
    \right|
    \ll
    \left|
        R^{(2)}_{\mu\nu}
    \right|\,.
    \label{DerivativeHierarchy}
\end{align}
We then neglect $R^{(1,1)}_{\mu\nu}$ while retaining the full metric dependence of $R^{(2)}_{\mu\nu}$.

This assumption should not be confused with an ordinary gradient expansion. If every derivative were assigned the same small parameter, $\partial g = O(\delta)$ would imply both $\partial^2g=O(\delta^2)$ and $(\partial g)^2=O(\delta^2)$, so that the two contributions to the curvature would generically appear at the same order. Equation~\eqref{DerivativeHierarchy} is instead an additional differential hierarchy, whose domain of validity must be verified for the solutions under consideration.

Importantly, the hierarchy~\eqref{DerivativeHierarchy} does not necessarily imply that the metric itself differs only slightly from a constant metric. Small local gradients may accumulate over an extended region and produce a finite or even order-unity variation of the metric coefficients. 

The distinction can be illustrated by a simple scalar analog of a metric component. Consider
\begin{align}
f(r) = 1 + a\ln\left(1+\frac{r}{r_0}\right)\,,
\end{align}
for which
\begin{align}
f'(r) = \frac{a}{r+r_0}\,, \qquad f''(r) = -\frac{a}{(r+r_0)^2}\,.
\end{align}
If $f$ is regarded schematically as a metric coefficient, the two structures entering the curvature are $f^{-1}f''$ and $(f^{-1}f')^2$. Their ratio is
\begin{align}
\frac{(f^{-1}f')^2}{|f^{-1}f''|}
=
\frac{|a|}
{\left|1+a\ln\left(1+r/r_0\right)\right|}\,.
\end{align}
Thus, for $|a| \ll 1$, the derivative-quadratic contribution can remain much smaller than the second-derivative contribution even when the metric coefficient itself is not close to its reference value. Indeed, for $r\gg r_0$,
\begin{align}
f(r) \simeq 1 + a\ln\frac{r}{r_0}\,,
\end{align}
and an order-unity accumulated variation is possible when $a\ln(r/r_0)=O(1)$. In this regime, the ratio above remains of order $|a|$, provided that $f$ stays away from zero. This example, therefore, explicitly shows how the derivative hierarchy may hold even when the metric cannot be treated globally as a small perturbation of a constant background.

The separation between $R^{(2)}_{\mu\nu}$ and $R^{(1,1)}_{\mu\nu}$ is not covariant: neither contribution is separately tensorial. The approximation is therefore defined, at this stage, in a chosen coordinate system. Whether a useful class of coordinates exists in which the hierarchy is preserved, and whether the resulting truncated equations are mutually consistent, are questions that must be addressed below.

Explicitly, the part of the Ricci tensor that is linear in the second derivatives of the metric is
\begin{align}
    R^{(2)}_{\mu\nu}
    =
    \frac{1}{2}g^{\alpha\beta}
    \left(
        g_{\beta\nu,\mu\alpha}
        +
        g_{\beta\mu,\nu\alpha}
        -
        g_{\mu\nu,\alpha\beta}
        -
        g_{\alpha\beta,\mu\nu}
    \right)\,.
    \label{RicciSecondDerivative}
\end{align}
The inverse metric in Eq.~\eqref{RicciSecondDerivative} has not been expanded. Consequently, even after the derivative-quadratic terms are discarded, the resulting equations remain nonlinear in the metric coefficients.

To make the derivative hierarchy dimensionally explicit, let $L$ denote a characteristic local length scale over which the ordering is
assessed, and introduce a dimensionless parameter $\delta \ll 1$. We assume schematically
\begin{align}
    g_{\mu\nu}
    &=O(1)\,,
    \\
    L\,\partial_i g_{\mu\nu}
    &=O(\delta)\,,
    \\
    L^2\,\partial_i\partial_j g_{\mu\nu}
    &=O(\delta)\,.
    \label{DerivativeOrdering}
\end{align}
It follows that
\begin{align}
    R^{(2)}_{\mu\nu}
    &=
    O\left(\frac{\delta}{L^2}\right)\,,
    &
    R^{(1,1)}_{\mu\nu}
    &=
    O\left(\frac{\delta^2}{L^2}\right)\,.
\end{align}
Thus, the terms quadratic in the first derivatives are parametrically smaller than those linear in the second derivatives.

For nonrelativistic matter, matching the retained curvature to the Einstein equations gives, schematically,
\begin{align}
    \frac{\delta}{L^2}
    \sim
    \frac{G\rho}{c^2}\,.
    \label{DerivativeMatterScaling}
\end{align}
Here, the large-$c$ limit is used only to organize the nonrelativistic ordering of the field and matter equations. Equation~\eqref{DerivativeMatterScaling} should therefore be understood as a scaling relation between the characteristic curvature and the matter source.

We next specify the nonrelativistic ordering of the mixed metric components. We restrict attention to coordinates adapted to the
nonrelativistic matter flow and assume
\begin{align}
    g_{0i}
    =
    \frac{\mathcal A_i}{c^2}
    +
    o(c^{-2})\,,
    \label{MixedMetricScaling}
\end{align}
while $g_{00}$ and $g_{ij}$ are retained at order unity and are not expanded around their Minkowski values. Consequently,
\begin{align}
    g^{00}
    =
    \frac{1}{g_{00}}
    +
    o(1)\,, \qquad
    g^{0i}
    =
    O(c^{-2})\,,
\end{align}
and the metric becomes block diagonal at leading nonrelativistic order.

More general large-$c$ scalings of the mixed sector may lead to additional nonrelativistic structures, as occurs in generalized Newton--Cartan limits.

Let
\begin{align}
    u^0
    =
    c\frac{dt}{d\tau}\,,
    \qquad
    u^i
    =
    v^i\frac{dt}{d\tau}\,.
\end{align}
The normalization condition
\begin{align}
    g_{\mu\nu}u^\mu u^\nu=-c^2
\end{align}
then gives
\begin{align}
    \frac{dt}{d\tau}
    =
    \frac{1}{\sqrt{-g_{00}}}
    +
    O(c^{-2})\,.
    \label{NonlinearTimeDilation}
\end{align}
Unlike in the weak-field limit, the leading relation between coordinate and proper time retains the full dependence on $g_{00}$.

For a perfect fluid, the leading components of the energy--momentum tensor are
\begin{align}
    T_{00}
    &=
    -\rho g_{00}c^2
    +
    O(c^0)\,,
    \label{NonlinearT00}
    \\
    T_{0i}
    &=
    -\rho c\,g_{ij}v^j
    +
    O(c^{-1})\,,
    \label{NonlinearT0i}
    \\
    T_{ij}
    &=
    \frac{\rho}{-g_{00}}
    g_{ik}g_{jl}v^kv^l
    +
    pg_{ij}
    +
    O(c^{-2})\,,
    \label{NonlinearTij}
\end{align}
while
\begin{align}
    T=-\rho c^2+O(c^0)\,.
    \label{NonlinearTrace}
\end{align}
We can now evaluate the derivative-truncated Einstein equations. At leading nonrelativistic order, the $00$ component of
Eq.~\eqref{RicciSecondDerivative} is
\begin{align}
    R_{00}^{(2)}
    =
    -\frac{1}{2}g^{ij}g_{00,ij}\,.
\end{align}
Using Eqs.~\eqref{NonlinearT00} and \eqref{NonlinearTrace}, the $00$ Einstein equation becomes
\begin{align}
    g^{ij}g_{00,ij}
    =
    \frac{8\pi G}{c^2}\rho g_{00}\,.
    \label{NonlinearEE00}
\end{align}
The spatial components are
\begin{align}
    R_{ij}^{(2)}
    ={}
    -\frac{1}{2}g^{00}g_{00,ij}
    +
    \frac{1}{2}g^{lm}
    \left(
        g_{mj,li}
        +
        g_{il,mj}
        -
        g_{ij,lm}
        -
        g_{lm,ij}
    \right)\,,
\end{align}
and therefore
\begin{align}
    -g^{00}g_{00,ij}
    +
    g^{lm}
    \left(
        g_{mj,li}
        +
        g_{il,mj}
        -
        g_{ij,lm}
        -
        g_{lm,ij}
    \right)
    =
    \frac{8\pi G}{c^2}\rho g_{ij}\,.
    \label{NonlinearEEij}
\end{align}
Here, the kinetic and pressure contributions to $T_{ij}$ are suppressed by $v^2/c^2$ and $p/(\rho c^2)$ relative to the rest-mass contribution entering through the trace.

Finally, the mixed sector is unsourced at the order retained here. In terms of the field $\mathcal A_i$ introduced in
Eq.~\eqref{MixedMetricScaling}, one obtains
\begin{align}
    g^{lm}
    \left(
        \mathcal A_{l,mi}
        -
        \mathcal A_{i,lm}
    \right)
    =0\,,
    \label{NonlinearEE0i}
\end{align}
up to terms of higher order in the large-$c$ and derivative hierarchies. We shall concentrate below on the block-diagonal sector,
while Eq.~\eqref{NonlinearEE0i} shows that a homogeneous Coriolis-like sector may, in principle, be retained.

For the matter sector, it is convenient to introduce
\begin{align}
    N\equiv\sqrt{-g_{00}}\,,
    \qquad
    \gamma_{ij}\equiv g_{ij}\,,
    \qquad
    \gamma\equiv\det\gamma_{ij}\,.
\end{align}
At leading nonrelativistic order,
\begin{align}
    u^0
    =
    \frac{c}{N}\,,
    \qquad
    u^i
    =
    \frac{v^i}{N}\,.
\end{align}
Since $\rho$ denotes the rest-mass density, we impose the conservation of the matter current,
\begin{align}
    \nabla_\mu(\rho u^\mu) = 0\,.
\end{align}
Using
\begin{align}
    \sqrt{-g}=N\sqrt{\gamma}
\end{align}
in the block-diagonal limit, the leading-order continuity equation is
\begin{align}
    \partial_t\left(\sqrt{\gamma}\rho\right)
    +
    \partial_i\left(\sqrt{\gamma}\rho v^i\right)
    =
    0\,.
    \label{NonlinearContinuity}
\end{align}
Thus, the nonperturbative lapse cancels from mass conservation, whereas the spatial metric enters through its volume element.

The spatial projection of energy--momentum conservation, $\nabla_\mu T^{\mu\nu}=0$, gives at leading nonrelativistic order
\begin{align}
    \left(
        \partial_t
        +
        v^j\partial_j
    \right)v^i
    -
    \frac{c^2}{2}
    \gamma^{ij}g_{00,j}
    -
    \frac{g_{00}}{\rho}
    \gamma^{ij}p_{,j}
    =
    0\,,
    \label{NonlinearEuler}
\end{align}
where terms involving the fluid velocity multiplied by the first derivatives of the spatial metric are subleading in the derivative and large-$c$ orderings adopted here.

Equivalently, using Eq.~\eqref{NonlinearContinuity}, Eq.~\eqref{NonlinearEuler} may be written in conservative form as
\begin{align}
    \partial_t
    \left(
        \sqrt{\gamma}\rho v^i
    \right)
    +
    \partial_j
    \left(
        \sqrt{\gamma}\rho v^i v^j
    \right)
    -
    \frac{c^2}{2}
    \sqrt{\gamma}\rho
    \gamma^{ij}g_{00,j}
    -
    \sqrt{\gamma}g_{00}
    \gamma^{ij}p_{,j}
    =
    0\,.
    \label{NonlinearEulerConservative}
\end{align}
As a consistency check, consider the weak-field limit ($|\Psi|/c^2 \ll 1$)
\begin{align}
    g_{00}
    =
    -\left(
        1+\frac{2\Psi}{c^2}
    \right)\,,
    \qquad
    \gamma_{ij}
    =
    \delta_{ij} +
O(c^{-2})\,.
\end{align}
Equation~\eqref{NonlinearEuler} then reduces to
\begin{align}
    \left(
        \partial_t+v^j\partial_j
    \right)v^i
    +
    \Psi_{,i}
    +
    \frac{1}{\rho}p_{,i}
    =
    0\,,
\end{align}
which is the standard Euler equation in a Newtonian gravitational field.

\subsection{Consistency of the derivative truncation}
\label{sec:bianchi}

An important consistency question concerns the differential identities satisfied by the truncated field equations. In the full theory, the contracted Bianchi identities,
\begin{align}
    \nabla^\mu G_{\mu\nu} = 0\,,
\end{align}
guarantee the compatibility of the Einstein equations with energy--momentum conservation.

To examine what remains of this property after the derivative truncation, let us decompose the Einstein tensor according to the same hierarchy used for the Ricci tensor,
\begin{align}
    G_{\mu\nu}
    =
    G^{(2)}_{\mu\nu}
    +
    G^{(1,1)}_{\mu\nu}\,,
\end{align}
where
\begin{align}
    G^{(2)}_{\mu\nu}
    \equiv
    R^{(2)}_{\mu\nu}
    -
    \frac{1}{2}g_{\mu\nu}R^{(2)}\,, \qquad
    G^{(1,1)}_{\mu\nu}
    \equiv
    R^{(1,1)}_{\mu\nu}
    -
    \frac{1}{2}g_{\mu\nu}R^{(1,1)}\,.
\end{align}
By construction,
\begin{align}
    G^{(2)}_{\mu\nu}
    =
    O\left(\frac{\delta}{L^2}\right)\,,
    \qquad
    G^{(1,1)}_{\mu\nu}
    =
    O\left(\frac{\delta^2}{L^2}\right)\,.
\end{align}
The exact Bianchi identity, then, implies
\begin{align}
    \nabla^\mu G^{(2)}_{\mu\nu}
    =
    -
    \nabla^\mu G^{(1,1)}_{\mu\nu}\,.
    \label{TruncatedBianchi}
\end{align}
Since $G^{(1,1)}_{\mu\nu}$ is quadratic in the first derivatives of the metric, its divergence schematically contains terms of the form
\begin{align}
    \partial\!\left[(\partial g)^2\right]
    \sim
    (\partial g)(\partial^2 g)\,,
    \qquad
    \Gamma(\partial g)^2
    \sim
    (\partial g)^3\,.
\end{align}
Under the ordering~\eqref{DerivativeOrdering}, these are respectively $O(\delta^2/L^3)$ and $O(\delta^3/L^3)$. Therefore,
\begin{align}
    \nabla^\mu G^{(2)}_{\mu\nu}
    =
    O\left(\frac{\delta^2}{L^3}\right)\,.
    \label{ApproximateBianchi}
\end{align}
The derivative-truncated Einstein tensor is thus divergence-free up to the order neglected in the approximation. In particular, the apparent third-derivative terms arising from the divergence of $G^{(2)}_{\mu\nu}$ cancel identically, as required by
Eq.~\eqref{TruncatedBianchi}; the remaining terms are products of the first and second derivatives of the metric and belong to the discarded order.

Consequently, if the truncated field equations are written as
\begin{align}
    G^{(2)}_{\mu\nu}
    =
    \frac{8\pi G}{c^4}T_{\mu\nu}
    +
    O\left(\frac{\delta^2}{L^2}\right)\,,
\end{align}
they imply
\begin{align}
    \frac{8\pi G}{c^4}\nabla^\mu T_{\mu\nu}
    =
    O\left(\frac{\delta^2}{L^3}\right)\,,
\end{align}
at the same level of approximation. Energy--momentum conservation is therefore compatible with the truncated gravitational equations to the retained order.

This result establishes the consistency of the truncation with the contracted Bianchi identities at leading order. It does not, however, remove its coordinate dependence: the separation between $G^{(2)}_{\mu\nu}$ and $G^{(1,1)}_{\mu\nu}$ is not itself covariant, and the class of coordinate transformations preserving the hierarchy remains to be specified.

\subsection{Coordinate dependence and harmonic gauge}
\label{sec:harmonic-gauge}

The derivative hierarchy introduced above is not invariant under an arbitrary coordinate transformation. This is unavoidable, since the separation of the Ricci tensor into terms containing second derivatives of the metric and terms quadratic in its first derivatives is not itself tensorial. The approximation should therefore be regarded as a gauge-fixed approximation rather than as a covariant truncation of the Einstein equations.

A particularly natural gauge for the present purpose is provided by harmonic coordinates,
\begin{align}
    \Gamma^\mu
    \equiv
    g^{\alpha\beta}\Gamma^\mu_{\alpha\beta}
    =
    0\,.
    \label{ExactHarmonicGauge}
\end{align}
In harmonic coordinates, the Ricci tensor can be written in the form
\begin{align}
    R_{\mu\nu}
    =
    -\frac{1}{2}
    g^{\alpha\beta}
    g_{\mu\nu,\alpha\beta}
    +
    Q_{\mu\nu}
    \left(
        g^{-1},
        \partial g,
        \partial g
    \right)\,,
    \label{HarmonicRicci}
\end{align}
where $Q_{\mu\nu}$ contains terms quadratic in the first derivatives of the metric. The approximation introduced in Sec.~\ref{sec:nonlinear} therefore amounts, in harmonic coordinates, to retaining the principal part of the reduced Einstein equations,
\begin{align}
    R_{\mu\nu}
    =
    -\frac{1}{2}
    g^{\alpha\beta}
    g_{\mu\nu,\alpha\beta}
    +
    O\left(\frac{\delta^2}{L^2}\right)\,.
    \label{TruncatedHarmonicRicci}
\end{align}
The metric coefficients appearing in the principal operator are kept fully nonlinear.

In the nonrelativistic ordering, derivatives with respect to $x^0=ct$ are subleading, and Eq.~\eqref{TruncatedHarmonicRicci}
reduces to
\begin{align}
    R_{\mu\nu}
    =
    -\frac{1}{2}
    g^{ij}g_{\mu\nu,ij}
    +
    O\left(\frac{\delta^2}{L^2}\right)\,,
\end{align}
at the order considered here. In the block-diagonal sector, the leading field equations consequently take the particularly simple form
\begin{align}
    g^{ij}g_{00,ij}
    &=
    \frac{8\pi G}{c^2}\rho g_{00}\,,
    \label{HarmonicNonlinear00}
    \\
    g^{lm}g_{ij,lm}
    &=
    -\frac{8\pi G}{c^2}\rho g_{ij}\,.
    \label{HarmonicNonlinearij}
\end{align}
The harmonic condition itself retains the nonlinear metric coefficients. Neglecting the mixed sector and terms containing time
derivatives, its spatial components give
\begin{align}
    g^{lm}
    \left(
        2g_{jl,m}
        -
        g_{lm,j}
    \right)
    =
    g^{00}g_{00,j}\,.
    \label{NonlinearHarmonicCondition}
\end{align}
In the weak-field limit, this reduces to the harmonic-gauge condition used in Sec.~\ref{Sec:Newtlim}.

The choice of harmonic coordinates does not eliminate the coordinate dependence of the approximation. Residual transformations between harmonic coordinate systems satisfy
\begin{align}
    \Box_g x'^\mu=0\,.
\end{align}
Such transformations do not necessarily preserve the derivative hierarchy. We therefore define as admissible those harmonic coordinate systems in which Eq.~\eqref{DerivativeOrdering} holds. Affine transformations provide a trivial example of transformations preserving this property, whereas a general characterization of the residual coordinate freedom compatible with the truncation lies beyond the scope of the present work.

The physical content of the approximation must ultimately be assessed through observables rather than through the values of individual
metric components. In particular, any effect that disappears under an admissible change of harmonic coordinates should be regarded as a coordinate artifact rather than a prediction of the truncated theory.

\subsection{Massive test-particle motion}

We next consider the motion of a slowly moving massive test particle. In the block-diagonal sector, at leading nonrelativistic order, the spatial geodesic equation is
\begin{align}
    \frac{d^2x^i}{dt^2}
    =
    -c^2\Gamma^i{}_{00}
    =
    \frac{c^2}{2}\gamma^{ij}g_{00,j}\,.
    \label{NonlinearMassiveAcceleration}
\end{align}
The omitted terms contain either additional powers of $v/c$ or time derivatives that are subleading in the nonrelativistic ordering. Introducing the lapse
\begin{align}
    N\equiv\sqrt{-g_{00}}\,,
\end{align}
Eq.~\eqref{NonlinearMassiveAcceleration} can equivalently be written as
\begin{align}
    \frac{d^2x^i}{dt^2}
    =
    -c^2N\gamma^{ij}N_{,j}\,.
    \label{NonlinearLapseAcceleration}
\end{align}
Thus, the acceleration depends not only on the spatial variation of the lapse but also on its local value. This dependence is retained because the metric coefficients themselves have not been expanded.

As a consistency check, in the weak-field regime, we have
\begin{align}
    N
    &=
    1+\frac{\Psi}{c^2}+\cdots\,,
    \\
    \gamma_{ij}
    &=
    \delta_{ij}
    +
    O(c^{-2})\,.
\end{align}
Equation~\eqref{NonlinearLapseAcceleration} then reduces, at leading order, to
\begin{align}
    \frac{d^2x^i}{dt^2}
    =
    -\delta^{ij}\Psi_{,j}\,,
\end{align}
which is the Newtonian equation of motion.

\subsection{Null geodesics}

Null trajectories differ qualitatively from the slow massive-particle limit considered above. Since the spatial coordinate velocity of a photon is of order $c$, terms proportional to $v^2$ that are suppressed for nonrelativistic matter contribute at the same order as the gravitational term involving $g_{00}$.

In the leading block-diagonal sector, neglecting time derivatives and the mixed metric components, the spatial geodesic equation is
\begin{align}
    \frac{d^2x^i}{dt^2}
    ={}&
    \frac{c^2}{2}
    \gamma^{ij}g_{00,j}
    -
    {}^{(3)}\Gamma^i{}_{jk}v^jv^k
    +
    g^{00}v^iv^j g_{00,j}\,,
    \label{NonlinearNullGeodesic}
\end{align}
where
\begin{align}
    {}^{(3)}\Gamma^i{}_{jk}
    =
    \frac{1}{2}\gamma^{il}
    \left(
        \gamma_{lj,k}
        +
        \gamma_{lk,j}
        -
        \gamma_{jk,l}
    \right)
\end{align}
is the Levi--Civita connection associated with the spatial metric $\gamma_{ij}$.

Writing $g_{00}=-N^2$, Eq.~\eqref{NonlinearNullGeodesic} becomes
\begin{align}
    \frac{d^2x^i}{dt^2}
    ={}&
    -c^2N\gamma^{ij}N_{,j}
    -
    {}^{(3)}\Gamma^i{}_{jk}v^jv^k
    +
    2v^iv^j\partial_j\ln N\,.
    \label{NonlinearNullLapse}
\end{align}
The null condition reads
\begin{align}
    \gamma_{ij}v^iv^j
    =
    N^2c^2\,.
    \label{NonlinearNullCondition}
\end{align}
The first term in Eq.~\eqref{NonlinearNullLapse} is the same gravitational contribution that governs the leading motion of massive
particles. The second term, however, is negligible in the nonrelativistic massive-particle limit but is of the same order for a
null trajectory since $v^2=O(c^2)$. Light propagation, therefore, probes the spatial metric directly, rather than only through its effect on the relation between the matter source and the lapse.

To make this distinction more explicit, define the local propagation direction by
\begin{align}
    n^i
    \equiv
    \frac{v^i}{Nc}\,,
    \qquad
    \gamma_{ij}n^in^j=1\,,
\end{align}
and the projector orthogonal to it,
\begin{align}
    P^i{}_j
    =
    \delta^i{}_j
    -
    n^in_j\,.
\end{align}
The last term in Eq.~\eqref{NonlinearNullLapse} is parallel to the propagation direction and therefore does not contribute directly to
the transverse coordinate acceleration. Projecting orthogonally to the ray gives
\begin{align}
    a_\perp^i
    =
    -c^2N
    P^i{}_j\gamma^{jk}N_{,k}
    -
    P^i{}_j
    {}^{(3)}\Gamma^j{}_{kl}v^kv^l\,.
    \label{NonlinearTransverseAcceleration}
\end{align}
In the weak-field limit, Eq.~\eqref{NonlinearNullGeodesic} reduces to the standard result discussed in Appendix~\ref{app:newtonian-details}. Away from that regime, however, the lapse and the spatial metric enter the null dynamics nonlinearly and need not combine in the same way as in the usual weak-field expansion. The derivative-truncated theory may therefore lead to light-deflection effects that differ from the standard weak-field prediction.

Equation~\eqref{NonlinearTransverseAcceleration} should nevertheless be regarded only as an indication of this possibility. Coordinate
accelerations are not observables, and any genuine departure from standard gravitational lensing must ultimately be established through a coordinate-independent quantity, such as a deflection angle defined with respect to physical observers or the corresponding optical observables.

\section{Illustrative phenomenological directions}
\label{sec:applications}

A detailed phenomenological analysis of the derivative-truncated theory is beyond the scope of the present work. Nevertheless, it is useful to indicate where departures from the standard weak-field description may
be expected to be most relevant. We consider here a simple static and spatially isotropic configuration, both as an illustration of the formalism and as a guide for future applications.

Let
\begin{align}
    ds^2
    =
    -A(\mathbf{x})c^2dt^2
    +
    B(\mathbf{x})\delta_{ij}dx^idx^j\,.
    \label{IsotropicMetric}
\end{align}
For the metric~\eqref{IsotropicMetric}, written in isotropic Cartesian spatial coordinates, one has
\begin{align}
    \sqrt{-g}
    &=
    \sqrt{AB^3}\,,
    &
    g^{ij}
    &=
    \frac{1}{B}\delta^{ij}\,.
\end{align}
The spatial harmonic-coordinate condition,
\begin{align}
    \partial_\nu
    \left(
        \sqrt{-g}\,g^{i\nu}
    \right)
    =0\,,
\end{align}
therefore gives
\begin{align}
    0
    &=
    \partial_j\left(\sqrt{-g}\,g^{ij}\right)
    =
    \partial_j\left(\sqrt{AB}\,\delta^{ij}\right)
    =
    \partial_i\sqrt{AB}\,.
\end{align}
Hence,
\begin{align}
    AB=\mathrm{const}\,.
\end{align}
For an asymptotically Minkowskian configuration, the integration constant can be normalized to unity, so that
\begin{align}
    AB = 1\,.
    \label{HarmonicAB}
\end{align}
The metric, therefore, contains a single independent function,
\begin{align}
    ds^2
    =
    -A(\mathbf{x})c^2dt^2
    +
    \frac{1}{A(\mathbf{x})}
    \delta_{ij}dx^idx^j\,.
    \label{HarmonicIsotropicMetric}
\end{align}
In this sector, the $00$ field equation \eqref{HarmonicNonlinear00} reduces to
\begin{align}
    \nabla^2 A
    =
    \frac{8\pi G}{c^2}\rho\,.
    \label{IsotropicAEquation}
\end{align}
The spatial field equation gives the same relation at the order retained in the derivative hierarchy. Indeed,
\begin{align}
    \nabla^2\left(A^{-1}\right)
    =
    -\frac{1}{A^2}\nabla^2A
    +
    \frac{2}{A^3}(\boldsymbol{\nabla}A)^2\,.
\end{align}
The last term belongs to the derivative-quadratic order, which is discarded in the present approximation.

\subsection{Slow massive-particle dynamics}

For a slowly moving massive particle, Eq.~\eqref{NonlinearMassiveAcceleration} becomes
\begin{align}
    \frac{d^2x^i}{dt^2}
    =
    -\frac{c^2}{2}
    A\,\delta^{ij}A_{,j}\,.
    \label{IsotropicMassiveAcceleration}
\end{align}
Thus, even though the field equation~\eqref{IsotropicAEquation} is linear in $A$, the relation between the metric function and the
particle acceleration remains nonlinear.

For a spherically symmetric configuration and a circular orbit,
Eq.~\eqref{IsotropicMassiveAcceleration} gives
\begin{align}
    \frac{V^2}{r}
    =
    \frac{c^2}{2}
    A(r)A'(r)\,,
    \label{CircularVelocityA}
\end{align}
where $V=r\,d\phi/dt$ is the coordinate tangential velocity in the isotropic coordinates of Eq.~\eqref{HarmonicIsotropicMetric}. The tangential velocity measured by a static observer is instead
\begin{align}
    V_{\rm phys}
    =
    \frac{V}{A}\,,
\end{align}
and therefore
\begin{align}
    V_{\rm phys}^2
    =
    \frac{c^2r}{2}\frac{A'}{A}
    =
    \frac{c^2r}{2}(\ln A)'\,.
    \label{PhysicalCircularVelocityA}
\end{align}
In the weak-field regime,
\begin{align}
    A
    =
    1+\frac{2\Psi}{c^2}+\cdots\,,
\end{align}
this reduces to the Newtonian relation
\begin{align}
    \frac{V_{\rm phys}^2}{r}
    =
    \Psi'(r)+\cdots\,.
\end{align}
Equation~\eqref{PhysicalCircularVelocityA} shows that the derivative-truncated theory can, in principle, modify the dynamics of slowly moving massive particles when the value of $A$ differs appreciably from unity. However, in systems in which the gravitational potential itself remains weak, the nonlinear corrections are correspondingly small. The simplest spherical configurations, therefore, do not suggest a large modification of galactic rotation velocities in the usual weak-potential regime. We shall not pursue this application further here.

\subsection{Light propagation}

Null trajectories appear more promising because they probe both the lapse and the spatial geometry at leading order. For a static metric, the spatial projections of null geodesics can equivalently be described as geodesics of the optical metric
\begin{align}
    d\ell_{\rm opt}^2
    =
    \frac{\gamma_{ij}}{N^2}
    dx^idx^j\,.
    \label{OpticalMetric}
\end{align}
For Eq.~\eqref{HarmonicIsotropicMetric}, where $N^2=A$ and $\gamma_{ij}=A^{-1}\delta_{ij}$, this becomes
\begin{align}
    d\ell_{\rm opt}^2
    =
    \frac{1}{A^2}
    \delta_{ij}dx^idx^j\,.
    \label{IsotropicOpticalMetric}
\end{align}
The corresponding effective refractive index is, therefore,
\begin{align}
    n_{\rm opt}
    =
    \frac{1}{A}\,.
    \label{OpticalIndex}
\end{align}
This result retains the full nonlinear dependence on the metric coefficient. In the weak-field regime,
\begin{align}
    n_{\rm opt}
    =
    1-\frac{2\Psi}{c^2}+\cdots\,,
\end{align}
and one recovers the standard weak-field gravitational deflection. For a weakly deflected ray, Eq.~\eqref{OpticalIndex} suggests
schematically
\begin{align}
    \boldsymbol{\alpha}
    \simeq
    -\int
    \boldsymbol{\nabla}_{\perp}\ln A\,
    d\ell\,,
    \label{SketchDeflection}
\end{align}
which reduces to
\begin{align}
    \boldsymbol{\alpha}
    \simeq
    -\frac{2}{c^2}
    \int
    \boldsymbol{\nabla}_{\perp}\Psi\,
    d\ell
\end{align}
in the ordinary weak-field limit.

Away from that regime, however, the quantity controlling light propagation is $\ln A$ rather than its linearized perturbation. This
makes gravitational lensing a particularly natural setting in which to search for observable consequences of the approximation. A proper analysis requires the construction of solutions satisfying the derivative hierarchy, together with a definition of the deflection angle in terms of the physical source, lens, and observer frames. Such an analysis is left for future work.

\section{Conclusions}
\label{sec:conclusions}

In this work, we have investigated a nonrelativistic approximation to general relativity in which the usual weak-field expansion of the
metric coefficients is not imposed. Instead, the approximation is based on a differential hierarchy: terms in the curvature containing
second derivatives of the metric are retained, whereas terms quadratic in its first derivatives are assumed to be parametrically smaller. The metric and inverse metric multiplying the retained derivatives are kept fully nonlinear.

This construction differs from both the standard Newtonian limit and systematic large-$c$ expansions based on Newton--Cartan geometry. It is not a covariant truncation of the Einstein equations, since the separation between second-derivative and derivative-quadratic terms is coordinate dependent. We have therefore formulated the approximation in harmonic coordinates, where the retained contribution coincides with the principal part of the reduced Einstein equations. The relevant domain is restricted to harmonic coordinate systems in which the assumed derivative hierarchy is satisfied.

The hierarchy itself does not require the metric coefficients to remain close to constant values. Small local variations can accumulate over extended regions, as illustrated by logarithmic profiles, while the quadratic derivative terms remain parametrically suppressed. A local length scale $L$ and a dimensionless parameter $\delta$ provide a convenient characterization of this regime,
\begin{align}
    L\,\partial g=O(\delta)\,,
    \qquad
    L^2\,\partial^2g=O(\delta)\,,
    \qquad
    \delta\ll1\,.
\end{align}
The retained and discarded contributions to the curvature are then of orders $\delta/L^2$ and $\delta^2/L^2$, respectively.

An important consistency check follows from the contracted Bianchi identities. Although the truncated Einstein tensor is not exactly
divergence-free, its divergence is of the same order as the terms discarded in the approximation. Energy--momentum conservation is therefore compatible with the truncated gravitational field equations to the retained order. This provides a nontrivial consistency condition for the derivative hierarchy.

In the nonrelativistic block-diagonal sector, the lapse and the spatial metric remain nonlinear. Slowly moving massive particles obey, at leading order,
\begin{align}
    \frac{d^2x^i}{dt^2}
    =
    -c^2N\gamma^{ij}N_{,j}\,,
\end{align}
which reduces to the Newtonian force law in the weak-field regime. Null trajectories are more sensitive to the nonlinear geometry:
spatial-connection terms that are negligible for slow massive matter contribute at leading order for light. This makes gravitational
lensing a particularly natural observable with which to investigate possible departures from the standard weak-field description.

As a simple illustration, for a static spatially isotropic metric in harmonic Cartesian coordinates, the harmonic condition implies that the temporal and spatial metric functions satisfy $AB=\mathrm{const}$. With asymptotically Minkowskian normalization, $AB=1$, so that the metric is determined by a single function. The corresponding optical metric depends nonlinearly on this function and suggests modifications of light propagation outside the weak-field regime. By contrast, the simplest spherical configurations indicate that corrections to the dynamics of slowly moving matter are small whenever the gravitational potential itself remains weak.

Another potentially interesting direction concerns gravitational-wave propagation. Before taking the nonrelativistic reduction, the derivative-truncated vacuum equations in harmonic coordinates retain
the quasilinear principal part
\begin{align}
    g^{\alpha\beta}
    \partial_\alpha\partial_\beta g_{\mu\nu}
    =0\,.
\end{align}
Thus, even after the derivative-quadratic terms have been discarded, the wave operator itself depends nonlinearly on the metric. The characteristic structure is consequently determined by the full metric rather than by a fixed Minkowski background. For ordinary weak oscillatory waves, however, the first nonlinear corrections of the form $h\,\partial^2 h$ are of the same amplitude order as the discarded terms $(\partial h)^2$, so that the present truncation cannot in general be used as a controlled approximation to nonlinear wave--wave interactions. More promising settings may include weak gravitational waves propagating on nonperturbative backgrounds, or special wave
geometries for which the derivative-quadratic terms are geometrically suppressed. We leave these possibilities for future investigation.

The present work is intended primarily to define and assess the approximation rather than to develop its phenomenology. Quantitative studies of gravitational lensing and gravitational-wave propagation, including solutions satisfying the derivative hierarchy and observables defined with respect to physical sources and observers, are natural next steps. It would also be useful to investigate more general geometries and to characterize more fully the residual harmonic-coordinate freedom compatible with the derivative ordering.

\begin{acknowledgments}
I am grateful to Mariateresa Crosta for bringing Ref.~\cite{Buchert:2023lxz} to my attention and for stimulating discussions on the Newtonian limit of general relativity.
\end{acknowledgments}

\appendix

\section{Details of the standard Newtonian limit}
\label{app:newtonian-details}

In this Appendix, we collect some details of the weak-field and large-$c$ expansion used in Sec.~\ref{Sec:Newtlim}. They are not needed for the main
argument of the paper but are useful for fixing conventions and for making explicit the approximations entering the standard Newtonian limit.

\subsection{Linearized field equations}

Consider the weak-field expansion
\begin{align}
    g_{\mu\nu}
    =
    \eta_{\mu\nu}
    +
    \epsilon h_{\mu\nu}
    +
    O(\epsilon^2)\,,
\end{align}
where $\epsilon$ is a formal expansion parameter. To first order,
\begin{align}
    g^{\mu\nu}
    =
    \eta^{\mu\nu}
    -
    \epsilon h^{\mu\nu}
    +
    O(\epsilon^2)\,,
\end{align}
and the Christoffel symbols are
\begin{align}
    \Gamma^\mu_{\rho\nu}
    =
    \frac{\epsilon}{2}
    \eta^{\mu\sigma}
    \left(
        h_{\sigma\rho,\nu}
        +
        h_{\sigma\nu,\rho}
        -
        h_{\rho\nu,\sigma}
    \right)
    +
    O(\epsilon^2)\,.
    \label{AppLinearChristoffel}
\end{align}
The Ricci tensor is therefore
\begin{align}
    R_{\mu\nu}^{(1)}
    =
    \frac{\epsilon}{2}
    \left(
        \partial_\alpha\partial_\mu h^\alpha{}_\nu
        +
        \partial_\alpha\partial_\nu h^\alpha{}_\mu
        -
        \Box h_{\mu\nu}
        -
        \partial_\mu\partial_\nu h
    \right)\,,
    \label{AppLinearRicci}
\end{align}
where
\begin{align}
    h
    \equiv
    \eta^{\mu\nu}h_{\mu\nu}
    =
    -h_{00}+h_{kk}\,.
\end{align}
Using the trace-reversed Einstein equations,
\begin{align}
    R_{\mu\nu}
    =
    \frac{8\pi G}{c^4}
    \left(
        T_{\mu\nu}
        -
        \frac{1}{2}g_{\mu\nu}T
    \right)\,,
\end{align}
and $x^0=ct$, the component equations at first order are
\begin{align}
    \epsilon
    \left(
        \nabla^2 h_{00}
        -
        \frac{2}{c}\dot h_{k0,k}
        +
        \frac{1}{c^2}\ddot h_{kk}
    \right)
    &=
    -\frac{16\pi G}{c^4}
    \left(
        T_{00}
        -
        \frac{1}{2}g_{00}T
    \right)\,,
    \label{AppLinEE00}
    \\
    \epsilon
    \left(
        \nabla^2h_{0i}
        -
        h_{0k,ik}
        +
        \frac{1}{c}\dot h_{kk,i}
        -
        \frac{1}{c}\dot h_{ki,k}
    \right)
    &=
    -\frac{16\pi G}{c^4}
    \left(
        T_{0i}
        -
        \frac{1}{2}g_{0i}T
    \right)\,,
    \label{AppLinEE0i}
    \\
    \epsilon
    \left(
        h_{00,ij}
        -
        \nabla^2h_{ij}
        +
        h_{k(i,j)k}
        -
        h_{kk,ij}
        +
        \frac{1}{c^2}\ddot h_{ij}
        -
        \frac{1}{c}\dot h_{0(i,j)}
    \right)
    &=
    \frac{16\pi G}{c^4}
    \left(
        T_{ij}
        -
        \frac{1}{2}g_{ij}T
    \right)\,.
    \label{AppLinEEij}
\end{align}
Repeated spatial indices are summed, and parentheses denote
symmetrization without a factor $1/2$.

\subsection{Energy--momentum conservation and test-particle motion}

The conservation law
\begin{align}
    \nabla_\mu T^{\mu\nu}=0
\end{align}
becomes, to first order in $\epsilon$,
\begin{align}
    \partial_\mu T^{\mu\nu}
    +
    \frac{\epsilon}{2}h_{,\rho}T^{\rho\nu}
    +
    \frac{\epsilon}{2}\eta^{\nu\sigma}
    \left(
        h_{\sigma\rho,\lambda}
        +
        h_{\sigma\lambda,\rho}
        -
        h_{\rho\lambda,\sigma}
    \right)
    T^{\rho\lambda}
    =
    O(\epsilon^2)\,.
    \label{AppLinConservation}
\end{align}
Its time and spatial components can be written as
\begin{align}
    T^{\mu0}{}_{,\mu}
    -
    \epsilon h_{0\rho,\sigma}T^{\rho\sigma}
    +
    \frac{\epsilon}{2c}
    \dot h_{\rho\sigma}T^{\rho\sigma}
    +
    \frac{\epsilon}{2}
    h_{,\rho}T^{\rho0}
    &=
    O(\epsilon^2)\,,
    \label{AppConservation0}
    \\
    T^{\mu i}{}_{,\mu}
    +
    \epsilon h_{i\rho,\sigma}T^{\rho\sigma}
    -
    \frac{\epsilon}{2}
    h_{\rho\sigma,i}T^{\rho\sigma}
    +
    \frac{\epsilon}{2}
    h_{,\rho}T^{\rho i}
    &=
    O(\epsilon^2)\,.
    \label{AppConservationi}
\end{align}
Test-particle motion follows from
\begin{align}
    \frac{d^2x^\mu}{d\lambda^2}
    +
    \Gamma^\mu_{\rho\nu}
    \frac{dx^\rho}{d\lambda}
    \frac{dx^\nu}{d\lambda}
    =0\,,
\end{align}
where $\lambda$ is an affine parameter. Inserting
Eq.~\eqref{AppLinearChristoffel} gives
\begin{align}
    \frac{d^2x^\mu}{d\lambda^2}
    =
    -\frac{\epsilon}{2}
    \eta^{\mu\sigma}
    \left(
        2h_{\sigma\rho,\nu}
        -
        h_{\rho\nu,\sigma}
    \right)
    \frac{dx^\rho}{d\lambda}
    \frac{dx^\nu}{d\lambda}
    +
    O(\epsilon^2)\,.
\end{align}
It is often convenient to use the coordinate time $t$ rather than an
affine parameter. Defining
\begin{align}
    v^i
    \equiv
    \frac{dx^i}{dt}\,,
\end{align}
and using $x^0=ct$, the spatial coordinate acceleration is
\begin{align}
    \frac{d^2x^i}{dt^2}
    =
    -\Gamma^i_{00}c^2
    -
    2\Gamma^i_{0j}cv^j
    -
    \Gamma^i_{jk}v^jv^k
    +
    \frac{v^i}{c}
    \left(
        \Gamma^0_{00}c^2
        +
        2\Gamma^0_{0j}cv^j
        +
        \Gamma^0_{jk}v^jv^k
    \right)\,.
    \label{AppCoordinateAcceleration}
\end{align}
Equation~\eqref{AppCoordinateAcceleration} is useful because it makes explicit which terms are suppressed in the slow-motion limit and which can survive for relativistic trajectories.

\subsection{Large-\texorpdfstring{$c$}{c} reduction}

For a perfect fluid,
\begin{align}
    T_{\mu\nu}
    =
    \left(
        \rho+\frac{p}{c^2}
    \right)
    u_\mu u_\nu
    +
    pg_{\mu\nu}\,.
\end{align}
Writing
\begin{align}
    u^0
    =
    c\frac{dt}{d\tau}\,,
    \qquad
    u^i
    =
    v^i\frac{dt}{d\tau}\,,
\end{align}
the normalization condition
\begin{align}
    g_{\mu\nu}u^\mu u^\nu=-c^2
\end{align}
gives, in the combined weak-field and nonrelativistic ordering,
\begin{align}
    \frac{dt}{d\tau}
    =
    1+O(c^{-2})\,.
\end{align}
Consequently,
\begin{align}
    T_{00}
    &=
    \rho c^2+O(c^0)\,,
    \\
    T_{0i}
    &=
    -\rho c v_i+O(c^{-1})\,,
    \\
    T_{ij}
    &=
    \rho v_i v_j
    +
    p\delta_{ij}
    +
    O(c^{-2})\,,
    \\
    T
    &=
    -\rho c^2+O(c^0)\,.
\end{align}
Taking
\begin{align}
    g_{\mu\nu}
    =
    \eta_{\mu\nu}
    +
    \frac{1}{c^2}h_{\mu\nu}
    +
    o(c^{-2})\,,
\end{align}
Eqs.~\eqref{AppLinEE00}--\eqref{AppLinEEij} reduce at leading order to
\begin{align}
    \nabla^2 h_{00}
    &=
    -8\pi G\rho\,,
    \\
    \nabla^2h_{0i}
    -
    h_{0k,ik}
    &=
    0\,,
    \\
    h_{00,ij}
    -
    \nabla^2h_{ij}
    +
    h_{k(i,j)k}
    -
    h_{kk,ij}
    &=
    8\pi G\rho\,\delta_{ij}\,.
\end{align}
Similarly, Eqs.~\eqref{AppConservation0} and \eqref{AppConservationi} give
\begin{align}
    \dot\rho
    +
    (\rho v_i)_{,i}
    &=
    0\,,
    \\
    (\rho v_i)^{\bullet}
    +
    \left(
        \rho v_i v_j
        +
        p\delta_{ij}
    \right)_{,j}
    -
    \frac{1}{2}\rho h_{00,i}
    &=
    0\,.
\end{align}
For a slowly moving test particle, Eq.~\eqref{AppCoordinateAcceleration}
reduces to
\begin{align}
    \frac{d^2x^i}{dt^2}
    =
    \frac{1}{2}h_{00,i}\,.
\end{align}
The identification $h_{00}=-2\Psi$ then reproduces the standard Newtonian equations.

\subsection{Null trajectories}

For null geodesics, the slow-motion approximation cannot be used because the spatial coordinate velocity is of order $c$. Terms
containing explicit inverse powers of $c$ in Eq.~\eqref{AppCoordinateAcceleration} may therefore contribute at
leading order.

For an isolated configuration, we may set the homogeneous $h_{0i}$ sector to zero and use
\begin{align}
    h_{ij}=h_{00}\delta_{ij}\,.
\end{align}
Keeping all terms that contribute at leading weak-field order then gives
\begin{align}
    \frac{d^2x^i}{dt^2}
    =
    h_{00,i}
    -
    \frac{2}{c^2}
    v^iv^k h_{00,k}\,.
    \label{AppPhotonAcceleration}
\end{align}
At zeroth order, the null condition implies
\begin{align}
    v^i=cn^i\,,
    \qquad
    n_i n^i=1\,.
\end{align}
The second term in Eq.~\eqref{AppPhotonAcceleration} is parallel to $n^i$ and therefore does not contribute to the transverse deflection. Introducing
\begin{align}
    P^i{}_j
    =
    \delta^i{}_j-n^in_j\,,
\end{align}
one obtains
\begin{align}
    a_\perp^i
    \equiv
    P^i{}_j
    \frac{d^2x^j}{dt^2}
    =
    P^i{}_j h_{00,j}\,.
\end{align}
Since $h_{00}=-2\Psi$,
\begin{align}
    a_\perp^i
    =
    -2P^i{}_j\Psi_{,j}\,.
\end{align}
The transverse gravitational acceleration of a light ray is therefore twice the value obtained by treating light as a Newtonian particle moving in the potential $\Psi$. This is the usual factor-of-two relativistic correction to the weak-field deflection of light.

\bibliographystyle{unsrt}
\bibliography{biblio}

\end{document}